\documentclass[prb,
twocolumn,
 amsmath,amssymb,
 aps,
]{revtex4-2}

\usepackage{graphicx}
\usepackage{dcolumn}
\usepackage{bm}
\usepackage{braket}

\usepackage{hyperref}
\hypersetup{colorlinks,urlcolor=blue,linkcolor=red,          
citecolor=blue,        
filecolor=blue}

\begin{document}

\title{Long-range and steady-state entanglement of nitrogen vacancy centers using microwaves as a drive and synthetic antiferromagnet as a  dissipator}

\author{Federico Garcia-Gaitan}
\author{Branislav K. Nikoli\'c}
\email{bnikolic@udel.edu}
\affiliation{Department of Physics and Astronomy, University of Delaware, Newark, DE 19716, USA}

\begin{abstract}
    The search for optimal schemes and dissipative environments that can mediate long-range entanglement between two distant nitrogen-vacancy centers (NVCs) in diamond has been the subject of vigorous efforts over the past decade due to potential applications of such microscopic solid-state qubits in quantum sensing and quantum computing. However, stabilizing entanglement of NVCs into a {\em steady state} poses a significant challenge, typically requiring tuning the environment {\em out of equilibrium}. Here we microscopically and rigorously  derive a Lindblad quantum master equation (QME) for a system of two {\em driven-dissipative} NVCs, where the drive is microwave radiation and dissipation is provided by a single magnetic bath that is kept in {\em equilibrium}. The jump operators within the Lindblad QME are expressed in terms of Keldysh Green's functions defined as   correlators of dipolar field operators of the bath. This QME makes it possible to predict precise conditions for long-range and steady-state entanglement of two NVCs, while suggesting that a synthetic antiferromagnet could be an {\em optimal} choice for a dissipative environment. By using realistic parameters from available experiments, we estimate steady-state concurrence reaching $\mathcal{C}\simeq 0.28$ for two NVCs separated by $\sim 100 \: \mathrm{nm}$.  
\end{abstract}

\maketitle

\section{Introduction}\label{sec:intro}

Optically addressable spin defects, such as nitrogen-vacancy  centers (NVCs) in diamond~\cite{Doherty2013},  are promising qubit platforms for applications in quantum sensing~\cite{Casola2018, Balasubramanian2008} and quantum computing~\cite{Bradley2019}. For example, exceptionally long coherence time of NVC qubits reaching \mbox{$\sim 1$ s}~\cite{Doherty2013, BarGill2013} could be used for quantum memories and networking~\cite{Bradley2019, Yao2013}. However, entangling NVCs at long distances is a key challenge because direct interaction between them is too weak for separations longer than \mbox{$\sim 10$} nm~\cite{Bradley2019, Rovny2025}. The {\em long-range} entanglement is required  because independent and simultaneous optical readout of two NVCs positioned at \mbox{$\lesssim 10$ nm} is virtually impossible due to the spatial and emission-wavelength overlap of their fluorescence. Thus, protocols for such entanglement have been intensely pursued over the past decade or so, as mediated by photons~\cite{Bernien2013}, phonons~\cite{Lemonde2018, Li2020d}, or magnons~\cite{Trifunovic2013,Fukami2021,Candido2021}. 

Additionally, NVCs are also amply utilized for quantum sensing of spintronic and magnonic phenomena~\cite{Casola2018,Solyom2018,Andrich2017,Finco2021,Ogawa2025} or exotic magnetic phases~\cite{Chatterjee2019} as they are highly sensitive to magnetic field noise sources~\cite{Rovny2024}. However, such quantum sensing schemes typically operate with a  single NVC~\cite{Wornle2021} or an ensemble of {\em unentangled} NVCs~\cite{Barry2024},  thereby limiting the temporal resolution~\cite{Herb2024}. Very recently, quantum sensing using entangled pairs of NVCs  has been achieved~\cite{Rovny2025}, but with limitations stemming from their small separation \mbox{$d \sim 6$ nm} and the need for a laser as an entangler. 


\begin{figure}[t!]
    \centering
    \includegraphics[width=1.0\linewidth]{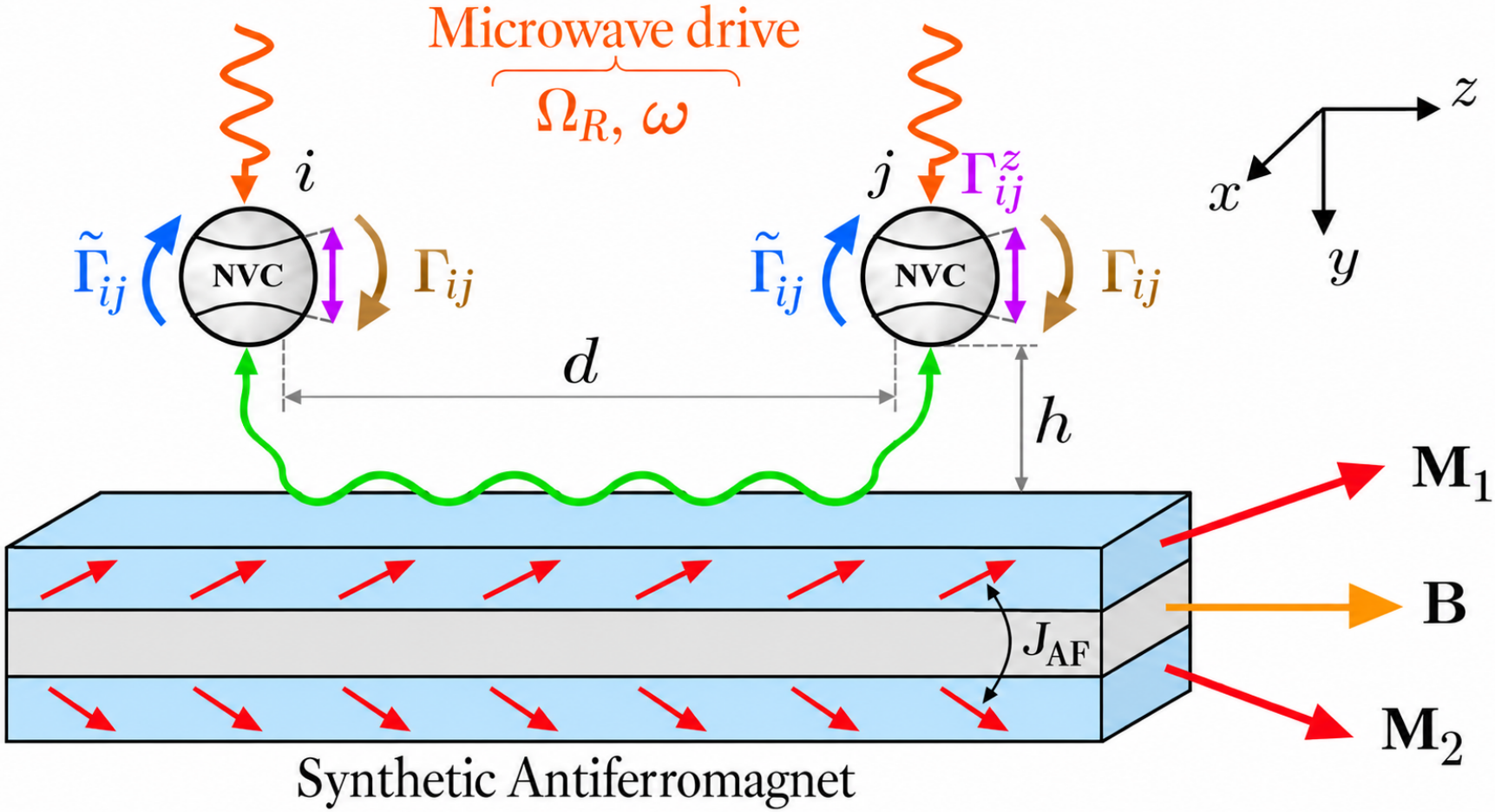}
    \caption{Schematic view of two NVCs (balls  separated by a distance $d$) driven by  microwaves (of frequency $\omega$ and amplitude imposing Rabi frequency~\cite{Doherty2013}  $\Omega_\mathrm{R}$) and dipolarly interacting (across distance $h$) with SAF acting as a dissipative bath.  The SAF, routinely fabricated in spintronics~\cite{Duine2018a}, consists of two ferromagnetic monolayers with antiferromagnetic  interlayer tunable exchange interaction $J_\mathrm{AF}$, where an external in-plane magnetic field $\mathbf{B}$ cants magnetizations $\mathbf{M}_1$ and $\mathbf{M}_2$ of two monolayers. Fluctuations of stray magnetic field due to excited magnons of the SAF couple two NVCs (as illustrated by the green wavy line) to induce {\em cooperative} (as NVCs interact with a shared single SAF bath) decay $\Gamma_{ij}$, pump $\tilde{\Gamma}_{ij}$, and dephasing $\Gamma^z_{ij}$ rates. The tunability of these parameters is enabled by the microwave drive,  thereby making possible {\em steady-state} entanglement of NVCs [Fig.~\ref{fig:entanglement}] while the SAF bath remains in {\em equilibrium}.}
    \label{fig:setup}
\end{figure}

While magnon-NVC coupling has been experimentally demonstrated~\cite{Andrich2017,Fukami2024, Candido2021}, typically by using magnons as excited states of standard magnetic insulator materials like YIG~\cite{Serga2010}, a clear-cut demonstration of magnon-mediated long-range entanglement between two distant NVCs is still lacking. 
The early analyses on how to entangle two NVCs at some height $h$ above magnetic material have relied on stray magnetic fields produced by its magnons~\cite{Trifunovic2013, Fukami2021, Candido2021, Neuman2020, GonzalesBallestero2022}. However, the thereby generated entanglement is only {\em transient}. This has motivated the search for schemes capable of producing {\em steady-state}  entanglement, as it could be achieved via topologically protected excitations~\cite{Hetenyi2022, Elman2017} or  stray fields produced by antiferromagnets~\cite{Flebus2019, Karanikolas2022}. Additionally, engineering  of dissipative environments could assist in achieving  steady-state entanglement of NVCs. 

Although dissipation has traditionally been considered as an impediment~\cite{Leggett1987}, causing decoherence and vanishing~\cite{GarciaGaitan2024,Parez2026} of entanglement,  recent experimental demonstrations~\cite{Liu2016,Spiecker2023,Harrington2022} 
of dissipation engineering have elevated it to an important resource~\cite{Verstraete2009} for  quantum information science. For example, in a system of superconducting qubits,  
a careful balance of driving and structured dissipation can produce controlled and resilient entanglement~\cite{Liu2016,Harrington2022, Zou2024}. A handful of schemes for engineering a magnetic dissipative bath to achieve steady-state entanglement of NVCs have been proposed   very recently~\cite{Zou2022,  Ullah2022, Xue2025, Dols2026,Kleinherbers2025,Nair2025, Hou2026, Xiong2022}, but they  typically require the bath to  be  finely tuned into a desired {\em nonequilibrium} state~\cite{Xiong2022, Hou2026, Nair2025}.



In this study, we predict that steady-state entanglement can be achieved even when the magnetic dissipative bath remains in thermal {\em equilibrium}. The key ingredient in our scheme [Fig.~\ref{fig:setup}] is microwave driving of two NVCs, which modifies the effective dissipative rates  in the dressed reference frame. Furthermore, it offers high tunability of entanglement,  and it can also be utilized to resolve small spectral features of magnetic bath, thereby opening new avenues for quantum sensing with more than one nontrivially coupled NVCs~\cite{Rovny2024, Rovny2025}. Our scheme emanates from the analysis of the Lindblad-type~\cite{Lindblad1976,Stefanini2026,Nathan2020,Manzano2020} quantum master equation (QME) that is microscopically derived [Eqs.~\eqref{eq:Lindblad}--\eqref{eq:Lindblad_detailed}; see also Appendix~\ref{sec:derivation}] for a driven-dissipative system of two distant NVCs interacting dipolarly [Fig.~\ref{fig:setup}] with a  magnetic bath positioned at a distance $h$ below them.  By proper tuning of the frequency of microwave driving and the bath spectral properties, we delineate  sufficient conditions for steady-state entanglement  persisting up to relatively long distances $d \sim 100\:\mathrm{nm}$ between two NVCs. Our theoretical analysis is applicable to various dissipative magnetic baths, and it can easily be applied to other optically addressable spin qubits. Nevertheless, we identify a synthetic antiferromagnetic (SAF)~\cite{Duine2018a} as an optimal bath, so that a system of NVCs + SAF is fully within the reach of present experimental capabilities~\cite{Jung2025, Finco2021, Duine2018a, Casola2018}. 

The paper is organized as follows. Sections~\ref{sec:models}   introduces Hamiltonian model  for two NVCs and a two-dimensional quantum magnet (2DQM), where the latter is traced over in Sec.~\ref{sec:lindblad} to arrive at Lindblad-type QME for Markovian dynamics of two NVCs viewed as an open quantum system. Results and discussion are provided in Sec.~\ref{sec:results}. We conclude in Sec.~\ref{sec:cando}. Additional Appendix~\ref{sec:derivation} provides further details of the Lindblad QME derivation; Appendix~\ref{sec:qme} provies derivation for a simpler case  where microwaves are absent; Appendix~\ref{sec:susc} provides derivation of the spin susceptibility tensor for the SAF~\cite{Duine2018a}, which can then be employed to express microscopically cooperative decay, pump, and dephasing rates entering the Lindblad QME; and Appendix~\ref{sec:deco} shows the effect in intrinsic decoherence of NVCs. 

\section{Models and Methods}\label{sec:mm}

\subsection{Hamiltonian models}\label{sec:models}

We consider [Fig.~\ref{fig:setup}] two identical NVCs separated by distance $d$  and placed at a height $h$ above a generic 2DQM, as it could be realized~\cite{Park2026} by quantum AF~\cite{Zhu2019c}, quantum ferromagnet, quantum spin liquid (QSL)~\cite{Broholm2020,Grover2013}, SAF composed of two quantum ferromagnets~\cite{Duine2018a}, etc. The total quantum system  NVCs + 2DQM is modeled by a Hamiltonian 
\begin{equation}\label{eq:totalhamiltonian}
\hat{H}=\hat{H}_\mathrm{NVCs}+\hat{H}_\mathrm{2DQM}+\hat{H}_\mathrm{int}.
\end{equation}
Here  
\begin{equation}\label{sec:hnvc}
\hat{H}_\mathrm{NVCs}=\sum_i \Delta \hat{\sigma}_i^z/2 + \Omega_\mathrm{R} \cos(\omega t)\hat{\sigma}_i^x, 
\end{equation}
describe two NVCs, $i \in \{1,2\}$, sufficiently far away (so that there is no direct interaction between them) and driven by microwaves of  frequency $\omega$ and  amplitude determining the Rabi frequency $\Omega_\mathrm{R}$  (i.e., the rate at which NVC flips between energy levels~\cite{Doherty2013}); $\hat{\sigma}_1^\alpha=\hat{\sigma}^\alpha \otimes \hat{I}_2$ and $\hat{\sigma}_2^\alpha=\hat{I}_2 \otimes \hat{\sigma}^\alpha$ are operators of two NVCs  obtained as the tensor product of  the Pauli matrices $\hat{\sigma}^\alpha$ and the unit $2 \times 2$ matrix $\hat{I}_2$; and we use $\hbar=1$ for simplicity. Thus, each  NVC is modeled as a two-level system with energy splitting $\Delta = \Delta_g-\gamma B$, where $\Delta_g =2.87\:\mathrm{GHz}$ is the zero-field splitting, $\gamma$ is the gyromagnetic ratio of the electron spin of the NVC, and $B$ is the magnitude of the externally  applied magnetic field. Note that $\Delta_g$ originates from the actual spin-1 nature of NVC~\cite{Doherty2013}, but the external magnetic field breaks the two-fold degeneracy of $|\! \pm 1\rangle$ states, thereby enabling easier description in terms of an effective spin-1/2 system that we adopt~\cite{Doherty2013}. Additional terms $\hat{H}_\mathrm{2DQM}$ and $\hat{H}_\mathrm{int}$ describe 2DQM below NVCs and the interaction between NVCs and 2DQM, respectively. A specific choice for $\hat{H}_\mathrm{2DQM}$ is given in Eq.~\eqref{eq:saf}.  We assume that $\hat{H}_\mathrm{int}$ originates from fluctuations in the dipolar stray field  
\begin{equation}\label{eq:hint}
    \hat{H}_\mathrm{int} = -\gamma \sum_i (\hat{B}_i^+\hat{\sigma}_i^-+\hat{B}_i^- \hat{\sigma}_i^+ + \hat{B}_i^z \hat{\sigma}_i^z),
\end{equation}    
where $\hat{B}_i^{\pm,z}$ is the corresponding dipolar magnetic field operator  and $\hat{\sigma}_i^\pm=(\hat{\sigma}_i^x \pm \hat{\sigma}_i^y)/2$. 

\subsection{Lindblad QME for a driven-dissipative system of two NVCs}\label{sec:lindblad}

We proceed to derive Lindblad QME for two NVCs by applying the unitary transformation 
\begin{equation}\label{eq:unitaryop1}
\hat{U} = \prod_i \exp (-i\omega t \hat{\sigma}_i^z/2),
\end{equation}
and by considering the rotating wave  approximation (RWA)~\cite{Doherty2013} to obtain the transformed Hamiltonian
\begin{equation}\label{eq:hrwa}
\hat{H}_\mathrm{NVCs}^\mathrm{RWA} \simeq \sum_i \delta
 \hat{\sigma}_i^z/2 + \Omega_\mathrm{R} \hat{\sigma}_i^x/2,
 \end{equation}
where the detuning \mbox{$\delta=\Delta-\omega$} can be minimized by choosing the driving frequency in resonance with the NVCs energy splitting. By rotating by an angle $\pi/2$ around the $y$-axis, via the unitary operator  
\begin{equation}\label{eq:unitaryop2}
\hat{U}_y=\prod_i\exp[-i(\pi/2)\hat{\sigma}_i^y/2]
\end{equation}
and by assuming negligible detuning, the interaction Hamiltonian becomes
\begin{widetext}
\begin{equation}
\label{eq:Hint_time}
\hat{H}_\mathrm{int} = -\gamma\sum_i \left\{-\hat{B}_i^z(t)[\hat{\sigma}_i^+e^{i\Omega_\mathrm{R} t} + \hat{\sigma}_i^- e^{-i\Omega_\mathrm{R} t}]+ \frac{\hat{B}^+_i(t)}{2} [\hat{\sigma}_i^z -\hat{\sigma}_i^+e^{i \Omega_\mathrm{R} t} + \hat{\sigma}_i^- e^{-i \Omega_\mathrm{R} t}]e^{-i\Delta t} + \frac{\hat{B}_i^-(t)}{2}  [\hat{\sigma}_i^z+\hat{\sigma}_i^+e^{i\Omega_\mathrm{R} t}-\hat{\sigma}_i^- e^{-i\Omega_\mathrm{R} t}]e^{i \Delta t} \right\}.
\end{equation}
\end{widetext}
Its explicit time-dependence arises from the interaction picture. 

A QME for the subsystem of two NVCs can be derived for this dressed interaction Hamiltonian as explained in Appendix~\ref{sec:derivation} (folloing similar steps as in Refs.~\cite{Li2025, Zou2022, Chatterjee2019}).  Such QME describing their  Markovian dynamics, as ensured by the weak coupling to 2DQM bath, has the standard Lindblad~\cite{Lindblad1976,Manzano2020,Nathan2020,Stefanini2026} form 
\begin{equation}\label{eq:Lindblad}
    \partial_t \hat{\rho} (t) = -i [\hat{H}_\mathrm{I},\hat{\rho}(t)] +\mathcal{L}[\hat{\rho}(t)],
\end{equation}
where  
\begin{equation}\label{eq:hi}
\hat{H}_\mathrm{I} = \sum_{i\neq j} (J_{ij} \hat{\sigma}_i^+ \hat{\sigma}_i^- +J_{ij}^z \hat{\sigma}_i^z \hat{\sigma}_i^z), 
\end{equation}
is the Hamiltonian accounting for how the traced-over-bath generates new effective  interactions  between NVCs, while bath-induced dissipation is captured by the Lindbladian $\mathcal{L}[\hat{\rho}(t)]$  
\begin{eqnarray}
    \label{eq:Lindblad_detailed}
    \mathcal{L}[\hat{\rho}(t)] &= \sum_{ij} \Gamma_{ij} \left(\hat{\sigma}_j^+\hat{\rho}(t)\hat{\sigma}_i^- -\frac{1}{2} \{\hat{\sigma}_i^- \hat{\sigma}_j^+ , \hat{\rho}(t) \} \right) \nonumber \\
    &+ \sum_{ij} \tilde{\Gamma}_{ij} \left ( \hat{\sigma}_j^- \hat{\rho} (t)\hat{\sigma}_i^+ -\frac{1}{2} \{\hat{\sigma}_i^+ \hat{\sigma}_j^-, \hat{\rho}(t) \}\right) \nonumber \\
    &+ \sum_{ij} \Gamma_{ij}^z \left(\hat{\sigma}_j^z \hat{\rho}(t)\hat{\sigma}_i^z -\frac{1}{2}\{\hat{\sigma}_i^z \hat{\sigma}_j^z, \hat{\rho}(t) \} \right).
\end{eqnarray}
Both coherent $J_{ij}$ and dissipative---$\Gamma_{ij}$, $\tilde{\Gamma}_{ij}$, and $\Gamma_{ij}^z$---parameters can be expressed (akin to Ref.~\cite{Li2025} where this was done in the absence of any drive) in terms of the Keldysh Green's functions (GFs)~\cite{Stefanucci2025} defined using dipolar field operators of 2DQM. For example, the coherent parameters  are expressed~\cite{Li2025} in terms of the retarded GF 
\begin{equation}\label{eq:gret}
G^R_{\hat{O}_1 \hat{O}_2} (\tau) = -i \Theta(\tau) \langle[\hat{O}_1(\tau),\hat{O}_2] \rangle, 
\end{equation}
with $\Theta$ being the Heaviside function,  as
\begin{subequations}\label{eq:jij}
    \begin{align}
        J_{ij} &= \gamma^2 \mathfrak{R}\left(G^R_{\hat{B}_i^z \hat{B}_j^z} (\Omega_\mathrm{R}) + \frac{1}{4} [ G^R_{\hat{B}_i^+ \hat{B}_j^-} (\nu_+) +G^R_{\hat{B}_i^+ \hat{B}_j^-} (\nu_-)] \right), \label{eq:J12}\\
        J_{ij}^z &= \frac{\gamma^2}{4}\mathfrak{R}[G^R_{\hat{B}_i^+ \hat{B}_j^-}(\Delta) + G^R_{\hat{B}_j^- \hat{B}_i^+}(\Delta)], \label{eq:J12z}
    \end{align}
\end{subequations}
where $\mathfrak{R} G^R_{\hat{O}_1 \hat{O}_2}(\Delta) = 1/2[G^R_{\hat{O}_1 \hat{O}_2}(\Delta) + G^R_{\hat{O}_2 \hat{O}_1}(-\Delta)]$, $\nu_\pm = \Delta \pm \Omega_\mathrm{R}$, and $\langle \ldots \rangle$ denotes the equilibrium expectation value. The dissipative parameters are expressed~\cite{Li2025} in terms of the lesser
\begin{equation}
    G^<_{\hat{O}_1 \hat{O}_2} (\tau)= -i \langle\hat{O}_1(\tau) \hat{O}_2 \rangle,
\end{equation}
and greater
\begin{equation}
G^>_{\hat{O}_1 \hat{O}_2} (\tau)= -i \langle\hat{O}_2 \hat{O}_1(\tau) \rangle,
\end{equation}
Keldysh GFs as  
\begin{subequations}\label{eq:gammafull}
\begin{align}
    \Gamma_{ij} &= i\gamma^2\left(G^>_{\hat{B}_i^z \hat{B}_j^z}(\Omega_\mathrm{R})+\frac{1}{4}[G^>_{\hat{B}_i^+ \hat{B}_j^-}(\nu_+) + G^<_{\hat{B}_i^+ \hat{B}_j^-}(\nu_-)] \right),  \label{eq:Gamma}\\
    \tilde{\Gamma}_{ij} &= i\gamma^2 \left(G_{\hat{B}_i^z \hat{B_j^z}}^<(\Omega_\mathrm{R}) +\frac{1}{4}[G^<_{\hat{B}_i^+ \hat{B}_j^-}(\nu_+) + G^>_{\hat{B}_i^+ \hat{B}_j^-}(\nu_-)]\right), \label{eq:Gammap}\\
    \Gamma_{ij}^z &= i\gamma^2 \left( G^<_{\hat{B}_i^+ \hat{B}_j^-}(\Delta) + G_{\hat{B}_i^+ \hat{B}_j^-}^>(\Delta) \right) \label{eq:Gammaz}.
    \end{align}
\end{subequations}
Equations ~\eqref{eq:jij} and ~\eqref{eq:gammafull} can also be recast in terms of the spin susceptibility tensor, as explicated in Appendix~\ref{sec:susc}. We note that a similar QME in the RWA appears~\cite{Hauss2008, Johansson2006} in the problem of a {\em single} driven superconducting qubit, although no analysis of its entanglement generation for a many-body system has been done hitherto.

Before delving into the entanglement generation mechanisms, we first discuss the limits of validity of the Lindblad QME we obtained in Eqs.~\eqref{eq:Lindblad}--\eqref{eq:Lindblad_detailed}. Besides the conventional weak system-bath coupling strength and short bath correlation times that are commonly assumed in the derivations of the Lindblad QME~\cite{Nathan2020,Stefanini2026}, Eqs.~\eqref{eq:Lindblad}--\eqref{eq:Lindblad_detailed} additionally require clear separation between the characteristic frequencies $\{\Omega_\mathrm{R}, \Delta, \nu_\pm \}$ and the decay rates in Eq.~\eqref{eq:gammafull}. For example, vanishing separation within any pair of characteristic frequencies would render the dissipative rate  separation in Eqs.~\eqref{eq:gammafull} spurious~\cite{Nathan2020}. Note that this requirement is satisfied in the NVCs + 2DQM system as the dissipative rates are in the range of \mbox{$\sim 100$ Hz} [Fig.~\ref{fig:SAF}(b)], whereas the characteristic Rabi and energy splitting frequencies for NVCs lie in the \mbox{$\sim 1$ MHz}  range. Finally, our QME could be further generalized to other systems and parameter regimes using strategies pursued in Refs.~\cite{Uchiyama2023, Vadimov2021, Shavit2019, Gulacsi2025}.


\begin{figure}[t!]
    \centering
    \includegraphics[width=1.0\linewidth]{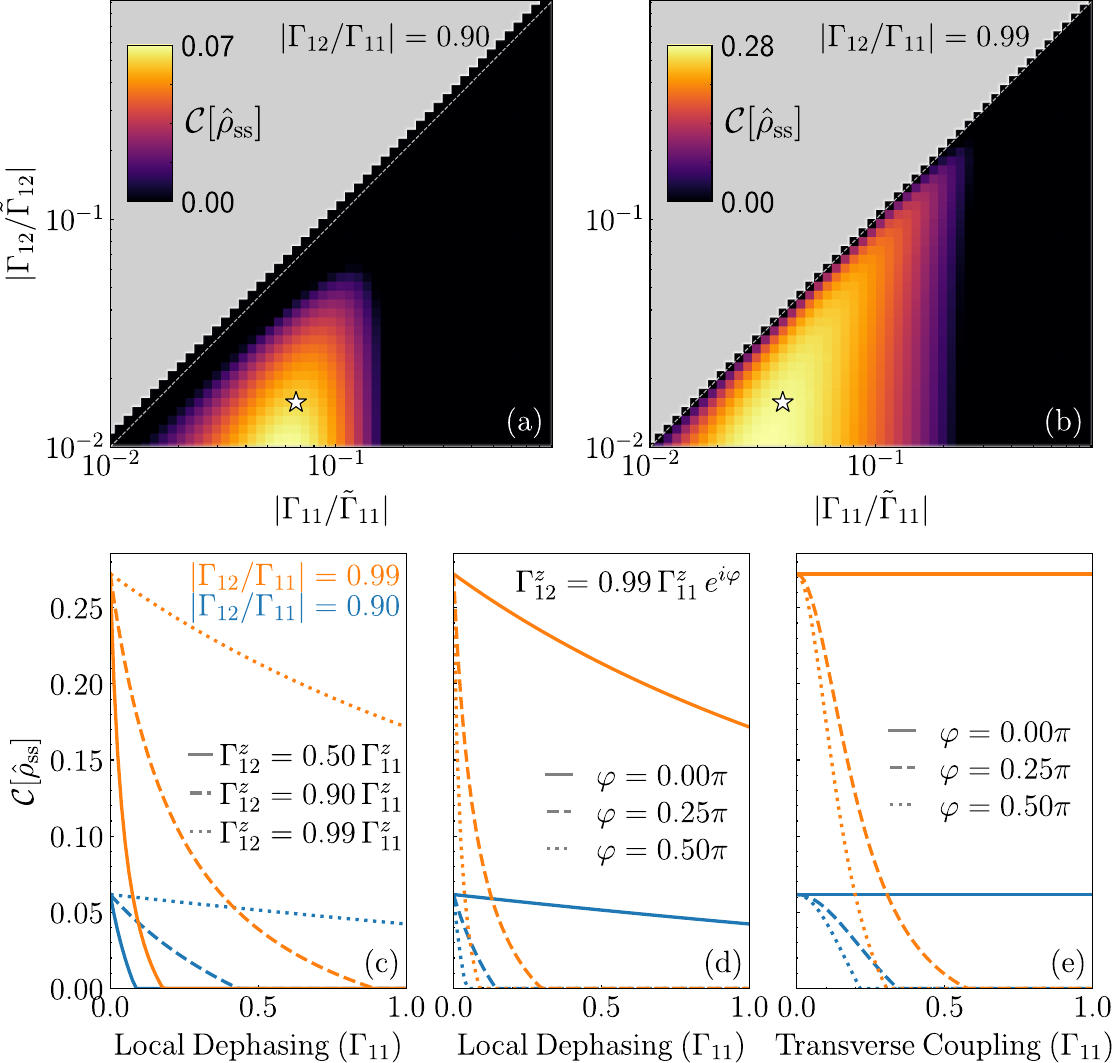}
    \caption{Steady-state entanglement of two NVCs in the setup of Fig.~\ref{fig:setup} with generic bath (not necessarily SAF), as quantified by the concurrence $\mathcal{C}[\hat{\rho}_{\mathrm{SS}}]$ of the steady-state density matrix [Eq.~\eqref{eq:rhoss}].  Panels (a) and (b) show $\mathcal{C}[\hat{\rho}_{\mathrm{SS}}]$ vs. the local and nonlocal pump-to-decay ratios $|\Gamma_{11}/\tilde{\Gamma}_{11}|$ and $|\Gamma_{12}/\tilde{\Gamma}_{12}|$  for the case in which the bath-induced dephasing [Eq.~\eqref{eq:Gammaz}] and coherent couplings [Eq.~\eqref{eq:jij}] are switched off. So, in this case, entanglement is generated solely by the cooperative decay [Eq.~\eqref{eq:Gamma}]  and pump [Eq.~\eqref{eq:Gammap}] rates, at fixed cooperativity $|\Gamma_{12}/\Gamma_{11}|=0.90$ in panel (a) and  $|\Gamma_{12}/\Gamma_{11}|=0.99$ in panel (b). The grey upper triangle in panels (a) and (b) denotes unphysical configurations. Panels (c)--(e) use the pump and decay rates marked by  stars in (a) and (b), corresponding to $|\Gamma_{12}/\Gamma_{11}|=0.90$ (blue) and $0.99$ (orange), to isolate the remaining entanglement-generation mechanisms relying on: (c) real cooperative dephasing $\Gamma^z_{12} \in \mathbb{R}$ and varying $\Gamma^z_{12}/\Gamma^z_{11}$, while $J_{ij}=J^z_{ij}=0$; (d) complex cooperative dephasing $\Gamma^z_{12}=0.99\,\Gamma^z_{11}e^{i\varphi} \in \mathbb{C}$; and (e) complex coherent  coupling  $J_{ij}\in \mathbb{C}$ [Eq.~\eqref{eq:J12}], while $\Gamma_{11}^z=\Gamma_{12}^z=0$.}
    \label{fig:entanglement}
\end{figure}

\begin{figure}
    \centering
    \includegraphics[width=0.99\linewidth]{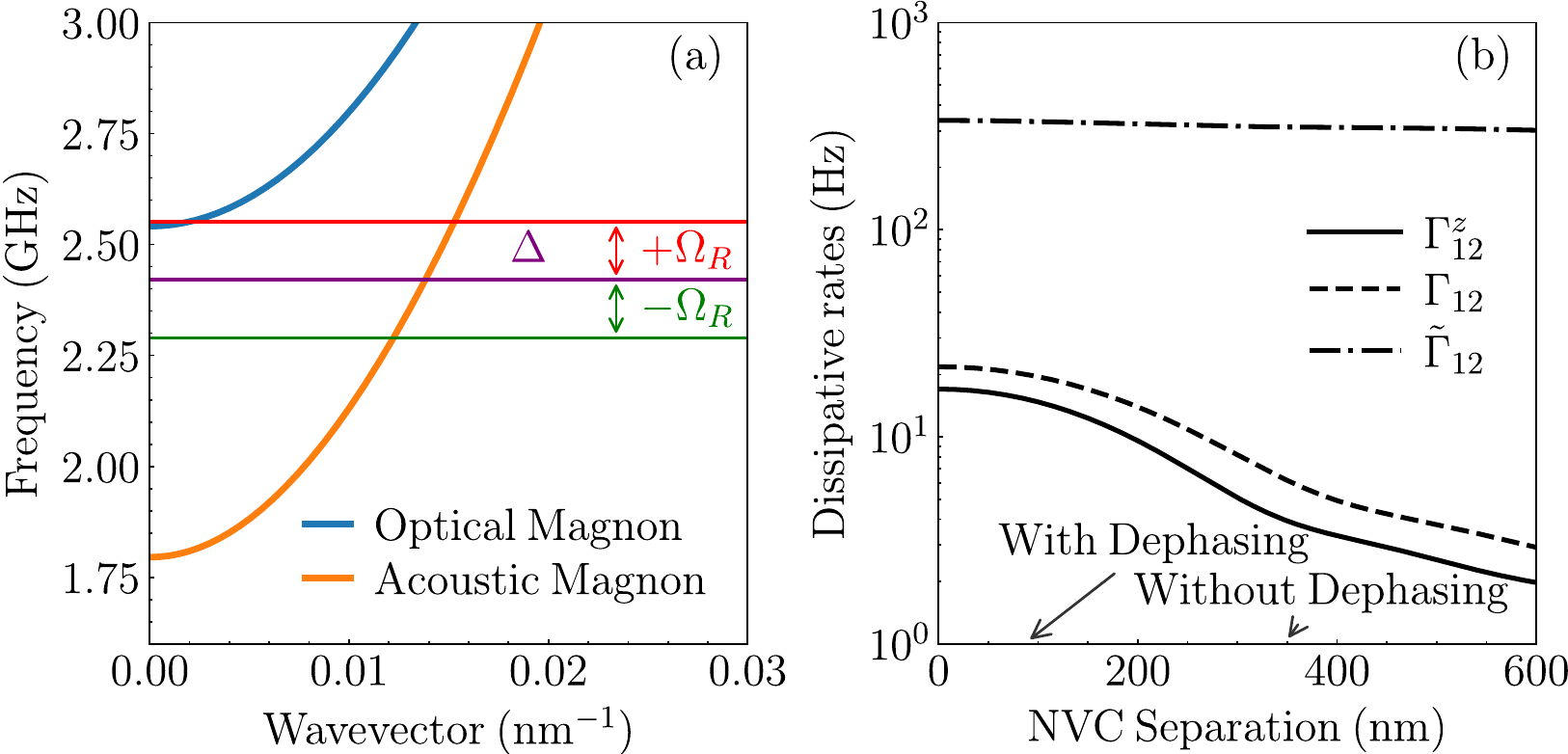}
    \caption{(a) Energy-momentum dispersion of magnons in SAF [Eq.~\eqref{eq:saf}], where the horizontal lines denote frequencies at which parameters for decay, pump, and dephasing rates  are evaluated [Eq.~\eqref{eq:gammafull}] in (b) as a function of the distance $d$ between two  NVCs. The maximum distances ensuring nonzero entanglement in Fig.~\ref{fig:entanglement}(a) are indicated by arrows at the abscissa. The shorter distance estimate is for the dephasing-induced constraint shown in Fig.~\ref{fig:entanglement}(c). The optimal position of the horizontal lines in (a) can be tuned by the Rabi frequency $\Omega_\mathrm{R}$ and energy splitting $\Delta$ of NVCs.}
    \label{fig:SAF}
\end{figure}

\section{Results and Discussion}\label{sec:results}

Let us first recall~\cite{Zou2022, Li2025} that in the absence of a drive, the quantum nature of the dissipative medium induces absorption $\Gamma_{ij}$ and emission $\tilde{\Gamma}_{ij}$ rates that are related via the fluctuation-dissipation theorem, \mbox{$\tilde{\Gamma}_{ij} = e^{-\Delta/T} \Gamma_{ij}$}, where $T$ is the temperature of the NVCs + 2DQM system. This genuinely quantum feature has been shown~\cite{Zou2022, Bellomo2015} to generate transient entanglement, but it leads to a steady state in the long time limit that is thermal Gibbs one and, therefore, unentangled. Conversely, QME [Eqs.~\eqref{eq:Lindblad}--\eqref{eq:Lindblad_detailed}] of driven NVCs possesses pump and decay rates that are no longer constrained by the fluctuation-dissipation theorem. This can easily be verified by taking the zero temperature limit of Eq.~\eqref{eq:Gammap} yielding  \mbox{$\tilde{\Gamma}_{ij}= i\gamma^2/4G^>_{\hat{B}_i^+ \hat{B}_j^-}(\nu_-)$} instead of $\tilde{\Gamma}_{ij}=0$ when the driving is absent. Thus, steady state will not converge into the Gibbs thermal state of the isolated  Hamiltonian of NVCs. A similar strategy was recently discussed in Ref.~\cite{Hou2026}, but requiring a  nonequilibrium bath to evade the Gibbs steady state. In contrast, our scheme evades Gibbs steady state by varying the Rabi frequency of NVCs determined by the amplitude of the microwave drive.

\subsection{General requirements for entanglement of two distant NVCs}\label{sec:generalreq}

Figure~\ref{fig:entanglement} analyzes the steady-state entanglement generation by exploring a wide range of possible dissipative parameters, while not being restricted by the properties of any specific bath choice. In it,  we compute concurrence $\mathcal{C}[\hat{\rho}_\mathrm{SS}]$~\cite{Wootters1998} from the steady-state (SS) density matrix solution $\hat{\rho}_\mathrm{SS}$ of our QME
\begin{equation}\label{eq:rhoss}
-i[\hat{H}_I,\hat{\rho}_\mathrm{SS}]+\mathcal{L}[\hat{\rho}_\mathrm{SS}]=0. 
\end{equation}
The concurrence takes values between $0$ and $1$, where entanglement is signaled by $\mathcal{C}[\hat{\rho}_\mathrm{SS}]>0$. In the absence of dephasing and coherent parameters, \mbox{$\Gamma_{ij}^z=J_{ij}=J_{ij}^z=0$},  the nonzero entanglement is determined~\cite{Hou2026} by tuning  the ratios of:  nonlocal and local decay rates, \mbox{$r=|\Gamma_{i\neq j}/\Gamma_{ii}|$}; local decay and  pump rates,  \mbox{$g_1=|\Gamma_{ii}/\tilde{\Gamma}_{ii}|$}; and nonlocal decay and pump rates,  \mbox{$g_2=|\Gamma_{i\neq j}/\tilde{\Gamma}_{i\neq j}|$}. For example, Fig.~\ref{fig:entanglement}(a),(b)  demonstrates  significant entanglement generation  for slow decay of the parameter $r$ and controlled asymmetry of the remaining parameters $g_1\neq g_2$. Figure~\ref{fig:entanglement}(c) highlights the role of the nonzero  dephasing rate $\Gamma_{ij}^z$ for two selected choices [indicated by stars in Fig.~\ref{fig:entanglement}(a),(b)] of  pump and decay rates. The intuitively detrimental role of dephasing on  entanglement, as observed by the quick decay [Fig.~\ref{fig:entanglement}(c),(d)] of $\mathcal{C}[\hat{\rho}_\mathrm{SS}]$ with increasing local dephasing rates, can {\em remarkably} be diminished by  sufficiently increasing nonlocal dephasing rates $\Gamma_{i\neq j}^z$. This feature can be viewed as  a realization of a decoherence-free subspace that has been amply studied in quantum information science~\cite{Lidar1998, Zanardi1997}. The additional role of local intrinsic decoherence is examined in Appendix~\ref{sec:deco}. Note, finally, that the advantage of large nonlocal dephasing rates is operative under the assumption of them being {\em real} [Fig.~\ref{fig:entanglement}(d)]. Conversely, bath-induced coherent parameters [Eq.~\eqref{eq:J12}] are detrimental for entanglement generation when $J_{i\neq j}$ are complex numbers  [Fig.~\ref{fig:entanglement}(e)], meaning that bath-induced Dzyaloshinskii–Moriya interaction~\cite{Zou2024} should be minimized. We also find bath-induced Ising interaction [Eq.~\eqref{eq:J12z}] to be  irrelevant  for the generation of entanglement of NVCs.

\subsection{General requriements for the properties of 2DQM bath}\label{sec:general}

Before discussing advantageous properties of SAF as a 2DQM bath, let us first  outline spectral properties of a generic 2DQM bath needed for steady-state entanglement generation in our scheme [Fig.~\ref{fig:setup}]. Firstly, the 2DQM should support long-range cooperative  pump $\tilde{\Gamma}_{ij}$  and decay rates $\Gamma_{ij}$, as they are the principal ingredient for entanglement generation in our scheme. Secondly, the cooperative dephasing rate $ \Gamma_{ij}^z$ must be carefully controlled [Fig.~\ref{fig:entanglement}(c)--(e)]---one strategy is to minimize those terms  by reducing the magnon spectral content of 2DQM in the range of NVCs energy splitting.  This option requires fine-tuning  beyond current experimental capabilities, such as achieving  significant magnon spectral content at both $\nu_\pm$ frequencies with simultaneous minimization of spectral content at $\Delta$. Alternatively, nonvanishing but slowly decaying dephasing rates [Fig.~\ref{fig:entanglement}(c)--(e)] would protect entanglement generation while making it easier for experimental realization.  Thirdly,  the Rabi frequency $\Omega_\mathrm{R}$ must be carefully tuned to resolve delicate spectral features of the 2DQM while still being significantly smaller than the  energy splitting of NVC. This  validates the RWA while enabling the tunability of parameters in the first and second  conditions. The  third condition  can be relaxed by considering driving that renders Hamiltonian $\hat{H}_\mathrm{NVCs}^\mathrm{RWA}$ exact, such as by using circularly polarized microwaves.

\subsection{Why choose SAF as a dissipative bath?}\label{sec:whysafbath}

As a specific 2DQM bath, we select SAF, routinely fabricated~\cite{Duine2018a} in spintronics by using two ferromagnetic layers with antiparallel magnetizations that are separated by a nonmagnetic metallic or insulating layer. This choice is motivated by  SAF being a highly tunable system where AF interlayer exchange interaction  can be varied by inserting~\cite{Duine2018a} a proper  interlayer of a nonmagnetic material. It also provides a minimum of two magnon bands {\em necessary} in our scheme. We describe SAF  by the following quantum Heisenberg Hamiltonian
\begin{eqnarray}\label{eq:saf}
    \hat{H}_\mathrm{SAF} &=& - J \sum_
    {\langle ij \rangle, L} \hat{\mathbf{S}}_{i,L} \cdot \hat{\mathbf{S}}_{j, L} +\sum_{i,L} \left[ \tilde{\gamma} \mathbf{B} \cdot\hat{\mathbf{S}}_{i,L}+K_z(\hat{S}_{i,L}^z)^2\right] \nonumber \\
    &+&J_{\mathrm{AF}} \sum_i \hat{\mathbf{S}}_{i, 1} \cdot \hat{\mathbf{S}}_{i, 2},
\end{eqnarray}
where  $\hat{\mathbf{S}}_{i,L}$ is the spin operator located at a site $i$ in the ferromagnetic layer $L=1,2$; $\langle ij \rangle$ indicates a sum over the nearest-neighbor sites; $J$ is the ferromagnetic intralayer exchange interaction; $J_\mathrm{AF}$ is the effective antiferromagnetic interlayer exchange interaction [Fig.~\ref{fig:setup}]; $K_z$ is an easy plane anisotropy; and $\tilde{\gamma} \mathbf{B}$ is the same external magnetic field applied to two NVCs, with $\tilde{\gamma}$ being the gyromagnetic ratio of local magnetic moments associated with $\hat{\mathbf{S}}_{i,L}$. Given the low NVC frequency operability range \mbox{$\sim 10 \: \mathrm{GHz}$} and the relatively large minimum height~\cite{Casola2018,Balasubramanian2008,Rovny2024,Rovny2025, Finco2021} \mbox{$h \geq 10 \: \mathrm{nm}$} between NVCs and 2DQM, an effective low-energy and long-wavelength approximation to $\hat{H}_\mathrm{SAF}$ [Eq.~\eqref{eq:saf}] is sufficient to derive the  Markovian QME of two NVCs after the SAF bath is traced over. Thus, the susceptibilities of the SAF entering into such QME are computed in Appendix~\ref{sec:susc} after the Holstein-Primakoff mapping~\cite{Primakoff1940} of  $\hat{\mathbf{S}}_{i,L}$ operators to bosonic ones and the linearization procedure~\cite{Gohlke2023,Bajpai2021,Chudnovsky2006} to arrive at the SAF Hamiltonian that is quadratic in bosonic operators. There we consider the SAF at zero temperature as entanglement mediated by it is enhanced~\cite{Li2025, Zou2022} at low temperatures.



Following three general requirements outlined in Sec.~\ref{sec:general}, an optimal configuration of SAF is illustrated in  Fig.~\ref{fig:SAF}. As discussed in Refs.~\cite{Hou2026, Li2025, Li2026}, the long-range nature of  cooperative dissipative couplings   [Eq.~\eqref{eq:gammafull}] can be enhanced by simultaneously considering larger interlayer stiffness, \mbox{$D=Ja^2=10.6$ meVnm$^2$} ($a$ is the lattice constant of each ferromagnetic layer of SAF), and by setting the NVC energy splitting close to the magnon band edges [in Fig.~\ref{fig:SAF}(a) we use  \mbox{$B=16$ mT} which causes \mbox{$\Delta=2.42$ GHz}]. The  height in Fig.~\ref{fig:setup} is set as \mbox{$h=100$ nm}, in accord with experiments~\cite{Finco2021}. The remaining SAF parameters are chosen as  \mbox{$J_\mathrm{AF}= 0.9$ GHz} and $K_z= 1.5 \:\mathrm{GHz}$ which guarantee that $\Delta$ lies between the acoustic and optical magnon bands [Fig.~\ref{fig:SAF}(a)]. Finally, the Rabi frequency is set to $\Omega_\mathrm{R} = 130\:\mathrm{MHz}$. While experimental implementation of large  $\Omega_\mathrm{R}$ is challenging, similar values have very recently been achieved~\cite{Jung2025}. Note also that large  $\Omega_\mathrm{R}$ simplifies the calculation of dissipative rates [Eq.~\eqref{eq:gammafull}] as there is no  spectral content at $\Omega_\mathrm{R}$ due to the magnon energy gap of SAF. The amplitude of microwave drive defining $\Omega_\mathrm{R}$ places the pump rate just above the acoustic magnon band, thus forcing it to  decay spatially slowly [Fig.~\ref{fig:SAF}(b)] while increasing its magnitude and keeping it within  \mbox{$\sim 100$ Hz}. These choices justify the usage of QME in Eqs.~\eqref{eq:Lindblad}--\eqref{eq:Lindblad_detailed}. Concurrently, cooperative dephasing and decay rates are also relatively long-ranged for the chosen intralayer exchange coupling $J$ [Fig.~\ref{fig:SAF}(b)]. Their lower magnitude allows us to locate ``sweet spots'' where conditions delineated in Fig.~\ref{fig:entanglement}(a) are easily satisfied. 

Note that detrimental effects from either dephasing or bath-induced complex coherent couplings reduce the distance between NVCs at which they can be long-range entangled in our scheme to $d \sim 100\:\mathrm{nm}$. By neglecting these effects, the distance can be spuriously increased to $d \lesssim 350\:\mathrm{nm}$. Finally, as shown in Appendix~\ref{sec:deco}, our approach is resilient against moderate intrinsic decoherence of NVCs. In particular, for the frequency rates indicates in Fig.~\ref{fig:SAF}(b), we predict our scheme to be robust for clean NVCs samples with coherence times \mbox{$\sim 1$ s}~\cite{Doherty2013, BarGill2013}. 



\section{Conclusions and Outlook}\label{sec:cando}

\subsection{Conclusions}\label{sec:conclusions}

The amply studied QMEs for a dissipative and not driven system of two NVCs, often using phenomenological Lindbladians~\cite{Zou2022},  have predicted dissipation-induced entanglement only as a transient phenomenon. Conversely, our microscopically  derived Lindblad QME [Eqs.~\eqref{eq:Lindblad}--\eqref{eq:Lindblad_detailed}] for the {\em driven-dissipative} system of two NVCs predicts the possibility of their steady-state entanglement. Our  QME no longer satisfies the detailed balance  between cooperative pump and decay rates, so it shares features of QMEs employed recently~\cite{Hou2026, Dols2026, Ullah2022}. However, unlike Refs.~\cite{Hou2026, Dols2026, Ullah2022} where both NVCs and the bath have to be driven, our scheme  requires 
{\em only driving} NVCs, which tunes their Rabi frequency, while the magnetic bath remains in {\em equilibrium}. This makes it possible to identify [Fig.~\ref{fig:entanglement}]  advantageous experimental conditions for the generation of steady-state entanglement of NVCs that have been overlooked in recent closely related studies~\cite{Hou2026, Dols2026}. Thus produced entanglement is robust to cooperative dephasing rates for carefully tuned parameters [Fig.~\ref{fig:entanglement}(c)--(e)]. In particular, we identify SAF~\cite{Duine2018a} as an optimal dissipative bath among possible choices of 2D quantum magnets, which could enabgle entanglement of two NVCs separated by \mbox{$d \sim 100$ nm} distance and with entanglement built up over a time scale of $\sim  0.1\:\mathrm{s}$. 

\subsection{Outlook}\label{sec:outlook}

It is worth mentioning that QSLs, which harbor native long-range entanglement~\cite{Grover2013} between their spins and are, therefore,  anticipated~\cite{Takei2024} (but also not yet fully confirmed in realistic materials~\cite{Broholm2020}) as potentially exceptional entanglers of distant NVCs, would actually not work in our scheme due to the short range of spin-spin correlations~\cite{GarciaGaitan2026} in QSLs. The long-range and steady-state entangled NVCs produced by our scheme could open new avenues for quantum sensing with more than one nontrivially coupled qubit~\cite{Rovny2024,Rovny2025}, which can then resolve fine spectral features in the excitation spectrum of magnetic baths. They can also be exploited as microscopic gates for highly scalable quantum computing hardware. 


\begin{acknowledgements}
This research was supported by the U.S. National Science Foundation (NSF) under Grant No. DMR-2500816. 
\end{acknowledgements}


\appendix

\section{Derivation of the Lindblad QME for two dissipative and driven NVCs}\label{sec:derivation}

Additional details of derivation leading to the Lindblad QME [Eqs.~\eqref{eq:Lindblad}--\eqref{eq:Lindblad_detailed}] are as follows. Starting from the von Neumann equation for time evolution of the density matrix of closed total system NVCs + 2DQM bain the interaction picture
\begin{equation} \label{eq:Neumanintpic}
    \partial_t \hat{\rho}_\mathrm{total}(t) = -i[\hat{H}_\mathrm{int}(t),\hat{\rho}_\mathrm{total}(t)],
\end{equation}
we follow the standard route~\cite{Breuer2007,Nathan2020,Stefanini2026} of tracing out bath degrees of freedom and applying Born and Markov approximations. This yields a time-local master equation of the Redfield type

\begin{equation} \label{eq:Redfield}
    \partial_t \hat{\rho}(t) = - \int_0^\infty\!\! d\tau \, \mathrm{Tr}_\mathrm{B}\left\{ \left[\hat{H}_\mathrm{int}(t),[\hat{H}_\mathrm{int}(t-\tau),\hat{\rho}(t)\otimes \hat{\rho}_\mathrm{B}]\right]\right\},
\end{equation}
where $\hat{\rho}_\mathrm{B}$ is the density matrix of the bath (which is assumed to be time independent and Gibbs-like), and $\mathrm{Tr}_\mathrm{B} \{\dots \}$ indicates the trace over the bath degrees of freedom. While Eq.~\eqref{eq:Redfield} is trace preserving, it can evolve the density matrix of the system toward losing its positive semidefinite properties~\cite{Breuer2007}. Thus, the secular approximation is commonly used to obtain a master equation in the Lindblad form by removing oscillating terms in Eq.~\eqref{eq:Redfield} explicitly depending on $t$. In our case, Eq.~\eqref{eq:Hint_time} can be conveniently rewritten as 
\begin{equation} \label{eq:Hint_simple}
\hat{H}_\mathrm{int}(t) = -\gamma\sum_{i}\sum_{\alpha, \beta \in \{ z,\pm\}} c_{\alpha,\beta} \hat{B}_i^\alpha (t) \hat{\sigma}_i^\beta e^{i(\nu_{\alpha} -\mu_\beta)t},
\end{equation}
where $\nu_\pm=\mp \Delta$, $\nu_z=\mu_z=0$, $\mu_\pm=\mp\Omega_\mathrm{R}$, are the frequencies appearing in the interaction picture in Eq.~\eqref{eq:Hint_time}, and $c_{\alpha,\beta}$ are simple prefactors tuned to match such equation. By plugging  Eq.~\eqref{eq:Hint_simple} into Eq.~\eqref{eq:Redfield}, we find that for each frequency appearing in Eq.~\eqref{eq:Hint_simple} only one partner term appears with the exact opposite frequency. By grouping these terms and splitting the double commutator structure into coherent and incoherent processes,  Eqs.~\eqref{eq:Lindblad}--\eqref{eq:Lindblad_detailed} are obtained. We stress that the secular approximation, employed to obtain the 
QME of the  Lindblad type, is valid as long as all frequencies are sufficiently separated~\cite{Nathan2020}.

\section{Derivation of the Lindblad QME for two dissipative but {\em not} driven NVCs}\label{sec:qme}

For a simpler Hamiltonian of two NVCs that does not include continuous microwave drive
\begin{equation}\label{eq:hamil}
\hat{H}_\mathrm{NVCs}=\sum_{i=1}^2 \Delta \hat{\sigma}_i^z/2.
\end{equation}
we trace over 2DQM bath to obtain reduced density matrix of NVCs alone~\cite{Li2025, Zou2022, Chatterjee2019} as a Lindblad   QME~\cite{Lindblad1976,Manzano2020,Stefanini2026,Nathan2020} given in Eqs.~\eqref{eq:Lindblad}--\eqref{eq:Lindblad_detailed}, but with simpler cooperative decay, pump and dephasing rates 
\begin{subequations}\label{eq:BathG}
\begin{align}
    \Gamma_{ij} &= i\gamma^2 G^>_{\hat{B}_i^+ \hat{B}_j^-}(\Delta),  \label{eq:GammaA}\\
    \tilde{\Gamma}_{ij} &= i\gamma^2 G_{\hat{B}_i^+ \hat{B}_j^-}^<(\Delta), \label{eq:GammapA}\\
    \Gamma_{ij}^z &= i\gamma^2G^>_{\hat{B}_i^z \hat{B}_j^z}(0)  \label{eq:GammazA},
    \end{align}
\end{subequations}
respectively. The fluctuation-dissipation theorem implies that $\tilde{\Gamma}_{ij}= \Gamma_{ij}e^{-\Delta/T}$, where $T$ is the temperature of the NVCs+2DQM total system. This constraint has been observed to render all dissipative-mediated entanglement only transient~\cite{Zou2022, Hou2026}. The 2DQM bath-induced coherent couplings from Eq.~\eqref{eq:jij} also simplify to
\begin{subequations}\label{eq:BathJ}
    \begin{align}
        J_{ij} &= \gamma^2 \mathfrak{R} G^R_{\hat{B}_i^z \hat{B}_j^z} (\Delta), \label{eq:J12A}\\
        J_{ij}^z &= \frac{\gamma^2}{4}\mathfrak{R} G^R_{\hat{B}_i^z \hat{B}_j^z}(0). \label{eq:J12zA}
    \end{align}
\end{subequations}
Additionally, evaluation of Eq.~\eqref{eq:BathJ} can be simplified by expressing the retarded GF in terms of the spin susceptibility tensor~\cite{Takei2024} 
\begin{equation}\label{eq:spinsusc}
G^R_{\hat{s}^\mu, \hat{s}^\nu}(\omega, \mathbf{q}) = -\chi^{\mu \nu} (\omega,\mathbf{q}). 
\end{equation}
Here we use the fact that dipolar field operators can be recast in terms of spin operators of the 2DQM bath~\cite{Li2025} as $\hat{\mathbf{B}}_i = \int d^2r \mathcal{D}_{\mathbf{r}_i, \mathbf{r}} \hat{\mathbf{s}}_\mathbf{r}$, where $\mathcal{D}_{\mathbf{r}_i,\mathbf{r}}$ is the classical Green's tensor of the dipolar magnetic field and $\hat{\mathbf{s}}_\mathbf{r}$ is the spin density operator. Thus, Keldysh GFs of dipolar operators can be expressed as
\begin{equation}
    G_{\hat{B}_i^\mu \hat{B}_j^\nu}^\lambda(\omega) = \int \frac{d^2\mathbf{k}}{(2\pi)^2} e^{i \mathbf{k}\cdot \mathbf{r}_{ij}}D_\mathbf{k}^{\mu \mu^\prime} D_{-\mathbf{k}}^{\nu\nu^\prime} G^\lambda_{\hat{s}^{\mu^\prime} \hat{s}^{\nu^\prime}}(\omega, \mathbf{k}).
\end{equation}
where  $\lambda \in \{<,>,R\}$; translational invariance of 2DQM and summation over repeated indices is assumed; and $D^{\mu \mu^\prime}_\mathbf{k}$ is the Fourier transform of the Green's tensor $\mathcal{D}_{\mathbf{r},\mathbf{r}^\prime}$. Furthermore, fluctuation-dissipation theorem makes it possible to recast Eq.~\eqref{eq:BathG} purely in terms of the spin susceptibility tensor as
\begin{widetext}
\begin{subequations}\label{eq:gamma}
\begin{align}
    \Gamma_{ij} &= 2\gamma^2 [n_\mathrm{B}(\Delta)+1] \int \frac{d^2\mathbf{k}}{(2\pi)^2} e^{i\mathbf{k}\cdot \mathbf{r}_{ij}} D^{+\mu}_\mathbf{k}D^{-\nu}_{-\mathbf{k}} \mathrm{Im}[\chi^{\mu\nu}(\Delta,\mathbf{k})],  \label{eq:Gammashort}\\
    \tilde{\Gamma}_{ij} &= 2\gamma^2 n_\mathrm{B}(\Delta) \int \frac{d^2\mathbf{k}}{(2\pi)^2} e^{i\mathbf{k}\cdot \mathbf{r}_{ij}} D^{+\mu}_\mathbf{k}D^{-\nu}_{-\mathbf{k}} \mathrm{Im}[\chi^{\mu\nu}(\Delta,\mathbf{k})], \label{eq:Gammapshort}\\
    \Gamma_{ij}^z &= \lim_{\omega \to 0^+}2\gamma^2 [n_\mathrm{B}(\omega)+1] \int \frac{d^2\mathbf{k}}{(2\pi)^2} e^{i\mathbf{k}\cdot \mathbf{r}_{ij}} D^{z\mu}_\mathbf{k}D^{z\nu}_{-\mathbf{k}} \mathrm{Im}[\chi^{\mu\nu}(\omega,\mathbf{k})],  \label{eq:Gammazshort}
    \end{align}
\end{subequations}
where $n_\mathrm{B}(\omega)$ is the Bose-Einstein function. Note that these expressions, also obtained in  Refs.~\cite{Chatterjee2019, Zou2022, Li2025}, are effectively utilized to compute dissipative rates in Eqs.~\eqref{eq:gammafull}.
\end{widetext}

\section{Spin susceptibility tensor of a synthetic antiferromagnet}\label{sec:susc}

By using the Holstein-Primakoff (HP)~\cite{Primakoff1940} bosonization of the SAF Hamiltonian in Eq.~\eqref{eq:saf}
\begin{subequations}\label{eq:hp}
    \begin{align}
        \hat{S}_{i,L}^z &= S-\hat{a}_{i,L}^\dagger\hat{a}_{i,L}, \label{eq:HPz}\\ 
        \hat{S}_{i,L}^- &= \sqrt{2S}\hat{a}_{i,L}^\dagger\sqrt{1-\frac{\hat{a}_{i,L}^\dagger \hat{a}_{i,L}}{2S}}, \label{eq:HP+}\\
        \hat{S}_{i,L}^+ &= \sqrt{2S}\sqrt{1-\frac{\hat{a}_{i,L}^\dagger \hat{a}_{i,L}}{2S}}\hat{a}_{i,L} \label{eq:HP-}, 
    \end{align}
\end{subequations}
where $\hat{a}_{i,L}^\dagger$ ($\hat{a}_{i,L}$) are bosonic creation (annihilation) operators, and by linearizing~\cite{Gohlke2023,Bajpai2021,Chudnovsky2006} these mappings, we obtain an analytical expression for the SAF  spin susceptibility tensor. For this purpose, we first determine the classical ground state, where the laboratory frame is  aligning the $z$-axis with the external magnetic field $\mathbf{B}$. We also assume that the $y$-axis  points out-of-plane [see Fig.~\ref{fig:setup}]. Then, the classical sublattice magnetizations are in the $xz$-plane
\begin{equation}
    \mathbf{M}_1 = \sin{\theta} \mathbf{x} + \cos{\theta}\mathbf{z}, \: \mathbf{M}_2= -\sin{\theta}\mathbf{x}+\cos{\theta}\mathbf{z},
    \label{eq:Classical_orientation}
\end{equation}
where the canting angle $\theta$ is determined by minimizing the classical energy. This quantity depends only on $|\mathbf{B}|$ and the antiferromagnetic Heisenberg exchange interaction between two ferromagnetic layers of SAF, \mbox{$\cos{\theta} = |\mathbf{B}|/2J_\mathrm{AF}S$}, 
where $S$ is the value of the spin is associated with operators $\hat{\mathbf{S}}_{i,L}$ in Eq.~\eqref{eq:saf}. Here the local coordinate axis of a spin in a layer $L=1,2$ is chosen so that the $z$-axis aligns with the directions specified in Eq.~\eqref{eq:Classical_orientation}. 

After Fourier transforming to momentum space and retaining only the quadratic terms~\cite{Gohlke2023,Bajpai2021,Chudnovsky2006} of bosonic operators,  the Hamiltonian can be rewritten in the Nambu basis $\Psi _\mathbf{k}=(\hat{a}_{\mathbf{k},1}, \hat{a}_{\mathbf{k},2}, \hat{a}_{-\mathbf{k},1}^\dagger, \hat{a}_{-\mathbf{k},2}^\dagger)^T$ as \mbox{$\hat{H}^{(2)}=\frac{1}{2}\sum_\mathbf{k} \Psi_\mathbf{k}^\dagger H_\mathbf{k}\Psi_\mathrm{k}$} with
\begin{widetext}
\begin{equation}
\label{eq:nambu}
\mathbf{H}_\mathbf{k} = \begin{pmatrix}
\epsilon_\mathbf{k} - C + K_d & A & K_a & B \\
A & \epsilon_\mathbf{k} -C + K_d & B & K_a \\
K_a & B & \epsilon_\mathbf{k} - C + K_d & A \\
B & K_a & A & \epsilon_\mathbf{k} - C + K_d
\end{pmatrix}.
\end{equation}
\end{widetext}
Here $\epsilon_\mathbf{k}=Dk^2+H\cos(\theta)$ is the magnon energy-momentum dispersion within each ferromagnetic layer of SAF; \mbox{$D=JSa^2$} is the spin-wave stiffness ($a$ is the lattice constant); $A=J_\mathrm{AF}S(1+\cos{2\theta})/2$, $B=J_\mathrm{AF}S(1-\cos{2\theta})/2$, and $C=A-B$ are terms produced by the interlayer antiferromagnetic exchange coupling of strenght $J_\mathrm{AF}$; and $K_d = -K_zS$ and $K_a=\frac{1}{2}K_zS$ are  terms due to magnetic anisotropy. Using additional bosonic operators
\begin{equation}
\hat{\alpha}_\mathbf{k}=\frac{\hat{a}_{\mathbf{k},1}+\hat{a}_{\mathbf{k},2}}{\sqrt{2}},\: \hat{\beta}_\mathbf{\mathbf{k}}= \frac{\hat{a}_{\mathbf{k},1}-\hat{a}_{\mathbf{k},2}}{\sqrt{2}},
\end{equation}
we block-diagonalize $H_\mathbf{k}$ into two independent Bogoliubov Hamiltonians,  $\hat{H}^{(2)}= \frac{1}{2}\sum_\mathbf{k} [\varphi_{\mathbf{k},1}^\dagger h_{\mathbf{k},1}\varphi_{\mathbf{k},1}+\varphi_{\mathbf{k},2}^\dagger h_{\mathbf{k},2}\varphi_{\mathbf{k},2}]$, where $\varphi_{\mathbf{k},1}=(\hat{\alpha}_\mathbf{k},\hat{\alpha}_{-\mathbf{k}}^\dagger)$, $\varphi_{\mathbf{k},2}=(\hat{\beta}_\mathbf{k},\hat{\beta}_{-\mathbf{k}}^\dagger)$, and 
\begin{equation}
h_{\mathbf{k},1(2)} = \begin{pmatrix} \ell_{\mathbf{k},1(2)} & \delta_{\mathbf{k},1(2)} \\ \delta_{\mathbf{k},1(2)} & \ell_{\mathbf{k},1(2)} \end{pmatrix}.
\end{equation}
Here the first block is a submatrix corresponding to the acoustic magnon band with its matrix elements given by
\begin{subequations}
    \begin{align}
        \ell_{\mathbf{k},1}&= \epsilon_\mathbf{k}+J_\mathrm{AF}S\sin^2{\theta}-K_zS,\\
        \delta_{\mathbf{k},2}&= J_\mathrm{AF}S \sin^2\theta + \frac{1}{2}K_zS,
    \end{align}
\end{subequations}
while the second one corresponds to the optical magnon band  with its matrix elements given by 
\begin{subequations}
    \begin{align}
        \ell_{\mathbf{k},2}&= \epsilon_\mathbf{k}-\frac{1}{2}J_\mathrm{AF}S(1+3\cos{2\theta})-K_zS,\\
        \delta_{\mathbf{k},2}&= -J_\mathrm{AF}S\sin^2{\theta}+\frac{1}{2}K_zS.
    \end{align}
\end{subequations}
Each of these block submatrices can be further diagonalized by the  Bogoliubov-Valatin transformation~\cite{Chudnovsky2006}, yielding magnon dispersions
\begin{equation}
    \omega_{1(2)}(\mathbf{k})= \sqrt{\ell_{\mathbf{k},1(2)}^2-\delta_{\mathbf{k},1(2)}^2},\label{eq:magnon_disp}
\end{equation}
and bosonic operators $\hat{\gamma}_{\mathbf{k},1(2)}$ that  completely and exactly diagonalize Hamiltonian $\hat{H}^{(2)}$
\begin{equation}\label{eq:diagonalham}
\hat{H}^{(2)}=\sum_\mathbf{k} \left[ \omega_{1}(\mathbf{k}) \hat{\gamma}_{\mathbf{k},1}^\dagger \hat{\gamma}_{\mathbf{k},1}  + \omega_{2}(\mathbf{k}) \hat{\gamma}_{\mathbf{k},2}^\dagger \hat{\gamma}_{\mathbf{k},2} \right].
\end{equation}
Their linear combination relates them to original HP bosonic operators in Eq.~\eqref{eq:hp}  
\begin{subequations}\label{eq:Bog}
    \begin{align}
        \hat{a}_{\mathbf{k},1} &= \frac{1}{\sqrt{2}}\sum_{L=1}^2 (u_L \hat{\gamma}_{\mathbf{k},L}+v_L \hat{\gamma}_{-\mathbf{k},L}),\label{eq:Bog1}\\
        \hat{a}_{\mathbf{k},2} &= \frac{1}{\sqrt{2}}\sum_{L=1}^2 s_L(u_L\hat{\gamma}_{\mathbf{k},L}+v_L\hat{\gamma}_{-\mathbf{k},L}),\label{eq:Bog2}
    \end{align}
\end{subequations}
where $s_{1(2)}=\pm1$; $u_L = \sqrt{\frac{\ell_{\mathbf{k},L}}{2\omega_L(\mathbf{k})}+\frac{1}{2}}$, and $v_L=\mathrm{sgn}(\delta_{\mathbf{k},L})\sqrt{\frac{\ell_{\mathbf{k},L}}{2\omega_L(\mathbf{k})}-\frac{1}{2}}$ ($\mathrm{sgn}$ is the sign function). 

To summarize, time-dependent spin susceptibility tensor can be computed---at the level of a quadratic $\hat{H}^{(2)}$ Hamiltonian capturing  noninteracting magnons [Eq.~\eqref{eq:diagonalham}] as low-energy  excitations of the full Hamiltonian of SAF in Eq.~\eqref{eq:saf}---by first mapping the original localized spin operators $\hat{\mathbf{S}}_{i,L}$ of SAF into HP bosonic operators [Eq.~\eqref{eq:hp}] and then by re-expressing them in terms of bosonic operators $\hat{\gamma}_{\mathbf{k},1(2)}$ via Eq.~\eqref{eq:Bog}. The time dependence of the spin susceptibility tensor then stems from
\begin{equation}\label{eq:timegamma}
    \hat{\gamma}_{\mathbf{k},1(2)}(t) = \hat{\gamma}_{\mathbf{k},1(2)}(0) e^{-i \omega_{1(2)}(\mathbf{k}) t}
\end{equation}
governed by the magnon energy-momentum dispersion in Eq.~\eqref{eq:magnon_disp}. Inserting the spin susceptibility tensor, Fourier transformed into the frequency domain, into Eq.~\eqref{eq:gamma} finally yields dissipative rates plotted in Fig.~\ref{fig:SAF}. Note that evaluation of Eq.~\eqref{eq:gamma} is facilitated  by employing the Green's tensor of a dipolar field generated by a thin plane of local magnetic moments~\cite{Guslienko2011}
\begin{widetext}
\begin{equation}
  D_\mathbf{k} = \frac{e^{-hk}}{2k}
  \begin{pmatrix}
    -(k_x+k)^2               & -k_z^2                     & i\sqrt{2}\,k_z(k+k_x) \\[2pt]
    -k_z^2                        & -(k-k_x)^2             & i\sqrt{2}\,k_z(k-k_x) \\[2pt]
    -i\sqrt{2}\,k_z(k+k_x)  & -i\sqrt{2}\,k_z(k-k_x) & -2k_z^2
  \end{pmatrix},
  \label{eq:DipolarT}
\end{equation}
\end{widetext}
where $h$ is the distance of NVCs from SAF and $k=|\mathbf{k}|$.

\section{Effects of intrinsic decoherence}\label{sec:deco}

To understand the impact of intrinsic decoherence of NVCs on our proposed scheme, we make the following substitution in Eqs.~\eqref{eq:Lindblad}--\eqref{eq:Lindblad_detailed} 
\begin{equation} \label{eq:intdeco}
    \Gamma_{ii}^z \mapsto \Gamma_{ii}^{z,\mathrm{SAF}} + \Gamma_{\mathrm{intrinsic}},
\end{equation}
where $\Gamma_{ii}^{z,\mathrm{SAF}}$ is the decoherence rate due to SAF dissipative bath and $\Gamma_{\mathrm{instrinsic}}$ is the additional intrinsic decoherence rate. We start dynamics from the unentangled initial state of two NVCs, $\hat{\rho}(t=0)=|\!\!\uparrow\uparrow\rangle \langle \uparrow \uparrow\!\!|$, and vary $\Gamma_\mathrm{intrinsic}$ from 0 to \mbox{$20$ Hz}. 
\begin{figure}
    \centering
    \includegraphics[scale=1.0]{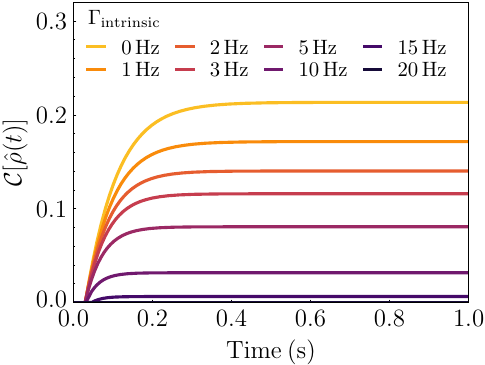}
    \caption{Time dependence of the concurrence of the  density matrix of two NVCs that is evolved by the QME in Eqs.~\eqref{eq:Lindblad}--\eqref{eq:Lindblad_detailed}, but with modified local decoherence rate [see Eq.~\eqref{eq:intdeco}].}
    \label{fig:deco}
\end{figure}
Figure~\ref{fig:deco} demonstrates the resilience of steady-state entanglement produced by our scheme to intrinsic decoherence of NVCs. This feature can also be understood from Fig.~\ref{fig:entanglement}(c), considering that local intrinsic decoherence is equivalent to reducing the ratio of nonlocal and local decoherence channels. Thus, intrinsic decoherence further restricts the total magnitude the net local decoherence frequency can reach before rendering steady-state concurrence negligible. For our scheme, this indicates that clean diamond samples hosting NVCs with long coherence times reaching \mbox{$\sim 1$ s}~\cite{Doherty2013, BarGill2013} could offer an ideal platform.

\bibliography{Biblio}

\end{document}